\documentstyle[aps,prc,preprint]{revtex}

\begin{document}
\title{Lattice Gauge Description of Colliding Nuclei}

\author{S. A. Bass, B. M\"uller, and W. P\"oschl}
\address{
Department of Physics, Duke University, Durham, NC 27708-0305, USA }

\date{\today}

\preprint{DUKE-TH-98-168}
\maketitle

\begin{abstract}
We propose a novel formalism for simultaneously describing both, the hard
and soft parton dynamics in ultrarelativistic collisions of nuclei.
The  emission of gluons from the initially coherent parton configurations 
of the colliding nuclei and low-$p_t$ color coherence effects are
treated in the framework of a Yang-Mills transport equation on a coupled
lattice-particle system. A collision term is added to the transport 
equation to account for the remaining intermediate and high-$p_t$ 
interactions in an infrared finite manner.
\end{abstract}
\bigskip

Experiments with heavy ion collisions at energies above
100 GeV/u, in preparation at the Relativistic Heavy Ion Collider (RHIC)
in Brookhaven and the Large Hadron Collider (LHC) at CERN, will try to
establish the existence of a new phase of nuclear matter, the quark-gluon 
plasma \cite{Harris.96}. One of the theoretical challenges in this context
is to develop a description on the basis of quantum chromodynamics (QCD)
of the processes that may lead to the formation of locally equilibrated
superdense matter in these nuclear reactions.

In recent years, most theoretical attempts at developing such a description
have been based on the idea that, at very high energy and for heavy nuclei,
the dominant mechanism of energy deposition in the central kinematical
region is the perturbative scattering of partons 
\cite{Hwa.86,Blaizot.87,Eskola.89,Gyulassy.91}. 
Because the interactions among gluons are stronger
than those involving quarks, this mechanism predicts an abundance of gluons
during the early equilibration phase \cite{Shuryak.92}. This concept can
be generalized into a theoretical framework, called the parton cascade
model \cite{Geiger.92}, which formulates thermalization as a transport 
process involving perturbative QCD excitations, i.e. quarks and gluons
\cite{Geiger.97}. The predictions of this formalism have been extensively
studied by means of numerical simulations \cite{Geiger.95,VNI}.

One of the problems inherent in this formulation concerns the description
of the initial state. The transport equations start with the assumption
of a probabilistic phase space distribution of partons, whereas in reality
the states of the colliding nuclei are described by coherent parton wave
functions. The incoherent parton description fails, especially, at small
transverse momenta, because the QCD coupling constant diverges in naive
perturbation theory. Some time ago it was proposed that the proper solution
to these difficulties would be the perturbative expansion, not around the
"empty" QCD vacuum, but around a mean color field describing the static
color field accompanying the fast moving valence quarks of the colliding
nuclei \cite{McLerran.94}. 

Because the mean color field of a heavy nucleus locally receives 
contributions from the quarks contained in many different nucleons, its 
source can be represented as a Gaussian ensemble of color charges moving 
along the light cone \cite{Kovchegov.96}. We will, therefore, refer to
this model here as the random light-cone source model (RLSM). Within
this framework, the energy deposition by gluonic interactions is described
as classical gluon radiation at small transverse momenta 
\cite{Kovner.95,Kovchegov.97}, and as gluon-gluon scattering at high
transverse momenta \cite{Gyulassy.97}. Quantum corrections to this picture
\cite{Jalilian.97} predict an enhancement of the glue field of the 
colliding nuclei at small values of the Bjorken variable $x$.
The full solution of the nonlinear classical RLSM equations for the 
color field of two colliding nuclei requires a lattice formulation
in 2+1 dimensions\cite{Krasnitz.97}.

The possibility of a description of inelastic gluon processes by means
of the nonlinear interactions of classical color fields has also been
explored numerically in studies of collision of two Yang-Mills field 
wave packets on a one dimensional gauge lattice \cite{Hu.95}. These
calculations gave evidence that the interaction between localized
classical gauge fields can lead to the excitation of long wavelength 
modes in the collision, which is reminiscent of the production of an
equilibrated gluon plasma.

Here we address the question how this new insight can be incorporated 
into the conceptual framework of the parton cascade model. First of all,
it is necessary to include a coherent color field $A_\mu$, in addition
to the incoherent quark and gluon distributions, $q_{\rm f}(r,p)$ and
$g(r,p)$. The subscript ``{f}'' here denotes the various quark flavors.
We will also insist on a full (3+1)-dimensional representation, which
will permit the study of deviations from boost invariance. 

Because even the classical Yang-Mills equations do not, in general, 
allow for global analytic solutions \cite{Matinyan.81,Froyland.83},
we propose to solve the RLSM equations numerically on 
a gauge lattice. Lattice calculations in Euclidean space-time 
have been shown to provide a reliable approach for the calculation 
of static and quasi-static properties of strongly coupled quantum 
field theory, in particular, QCD. For dynamical systems far off
equilibrium, however, one needs to study the system in real
continuous time. The lattice discretization then should only be
applied to the Euclidean sub-space ${\bf R^3}$. In this case
it is appropriate to choose a Hamiltonian formulation rather than
a Lagrangian one. We have to emphasize that this concept is neither 
explicitly invariant under general gauge transformations nor
Lorentz invariant. However, we believe that for the type of problems 
described above this method is indeed useful.

One has to select a rest frame in the space ${\bf R}\otimes {\bf R^3}$ 
which in our case probably is best chosen as the center of velocity. 
Further, one has to adopt a gauge. The temporal gauge in the continuum 
($A^0 = 0$) seems most appropriate here \cite{Kogut.75}. 
A set of equations describing the evolution of the phase space
distribution of quarks and gluons in the presence of a mean color field,
but in the absence of collisions, was proposed more than a decade ago by 
Heinz \cite{Heinz.83,Elze.89}. This non-Abelian generalization of the 
Vlasov equation can be considered as the continuum version of the dynamics
of an ensemble of classical point particles endowed with color charge
and interacting with a mean color field. The equations for this dynamical
system were originally derived by Wong \cite{Wong.70}.  

In the following we develop a formulation of the RLSM including the ideas 
of Heinz and Wong. We represent the valence quarks of the two 
colliding nuclei as point particles moving in the space-time continuum,
and interacting with a classical gauge field defined on a spatial lattice
but with quasi-continuous time.\footnote{The numerical implementation
also requires a discretization of the time variable, but the temporal step
size can be taken arbitrarily small.}  In principle, this idea follows 
the proposal of Hu and M\"uller \cite{Hu.97} for the simulation of the 
effects of hard thermal loops by means of colored point particles.

At this stage we are still general enough 
to assume that the soft modes of the gluon fields
are described through gauge fields with SU($N$) symmetry.
In the associated Lie-algebra LSU($N$) we express the
Hamiltonian of the above outlined system in the continuum as
\begin{eqnarray}
\label{Eq.1}
{H} & =& 
\sum\limits_{i=1}^{N_1}\sqrt{\vert{\vec p}_i\vert^2+m_0^2}\, +\,
\sum\limits_{i=N_1+1}^{N_2}\sqrt{\vert{\vec p}_i\vert^2+m_0^2} 
\nonumber \\
&-&2g\int d^3x\, {\rm Tr} [{\cal J}_{\mu}{\cal A}^{\mu}] 
\,-\,{1\over 2}\int d^3x\, {\rm Tr} [{\cal F}^{\mu\nu}{\cal F}_{\mu\nu}]
\end{eqnarray}
where ${\cal F}^{\mu\nu},{\cal A}^{\mu},{\cal J}^{\mu},
{\cal Q}_i \,\epsilon$ LSU($N$). 
The curly quantities denote those in the adjoint representation, which
are defined as, e.g., ${\cal A}^{\mu} = A^{\mu}_c\cdot T^c$ with group 
generators $T^c$.   $g$ is the gauge coupling constant, 
and ${\cal F}^{\mu\nu}$ denotes the 
field strength tensor of the mean color field ${\cal A}_{\mu}$. 
The moving particles generate a color current
\begin{equation}
\label{Eq.2}
{\cal J}^{\nu}(x) = \sum_i {\cal Q}_i(t) 
{{p^{\nu}_i}\over{\sqrt{\vert\vec p_i\vert^2+m_0^2}}} 
\delta(\vec x -\vec x_i(t)),
\end{equation}
where $t$ denotes the global time in the chosen reference frame.
Denoting the space-time positions, momenta, and color charges of the 
particles by $x_i^{\mu}$, $p_i^{\mu}$ and ${\cal Q}_i^a$, respectively, 
the following equations of motion are derived from the 
above Hamiltonian (\ref{Eq.1}):
\begin{eqnarray}
\label{Eq.3}
p_i^0{dx_i^{\mu} \over dt} &=& p_i^{\mu} \\
\label{Eq.4}
p_i^0{dp_i^{\mu} \over dt} &=& 2g\,{\rm Tr}({\cal Q}_i {\cal F}^{\mu\nu})
                                \, p_{i,\nu} \\
\label{Eq.5}
p_i^0{d{\cal Q}_i \over dt} &=& \imath g \, 
\left[ {\cal Q}_i ,{\cal A}^{\mu} \right]_- \, p_{\mu,i} 
\end{eqnarray}
The factors $p_i^0$ on the l.h.s. are needed to convert the derivatives
with respect to proper time into coordinate time derivatives.
Furthermore, the inhomogeneous Yang-Mills equations
\begin{equation}
D_{\mu}{\cal F}^{\mu\nu}(x)\, =\, g\, {\cal J}^{\nu}(x)
\label{Eq.6}
\end{equation}
describe the dynamics of the classical mean color fields.
The current density (\ref{Eq.2}) forms the source term on 
the r.h.s. of (\ref{Eq.6}).  The coupled system of the Wong equations
(\ref{Eq.3}--\ref{Eq.5}) and the Yang-Mills equation (\ref{Eq.6}) is 
highly nonlinear and can only be solved numerically or perturbatively.

These equations have been used to simulate the effects of hard
thermal loops \cite{Pisarski.89} on the dynamics of soft modes of a
non-Abelian SU(2) gauge field at finite temperature \cite{Hu.97,Moore.98}.
In this case, the colored particles describe the gauge field modes with
thermal momenta, and the mean field describes the coherent motion of those
gauge field modes which have a wave number $k$ much smaller than the
temperature $T$ and are highly occupied. The separation of the two regimes
was achieved by discretizing the mean gauge field on a lattice with elementary
spacing $a\ll T^{-1}$. Requiring particles to have momenta $p > \pi/a$ then
avoids double counting degrees of freedom.

Here we propose to use the equations (\ref{Eq.2}--\ref{Eq.6}) to describe 
the interactions among the glue field components of two colliding heavy 
nuclei. In this case, the lattice cut-off $a$ can be used to separate the
regime in transverse momentum where the dynamics of gluons is perturbative
(large $k_{\rm T}$) from that where naive perturbation theory fails
(small $k_{\rm T}$). The gluon propagators used for the calculation of
the collision terms will be regulated in the infrared by the lattice
cut-off $k_{\rm c} = \pi/a$. The interaction with the mean color field
allows for an exchange of an arbitrary number of gluons, and the screening
of the soft components of the gauge field by perturbative partons
\cite{Biro.92,Eskola.96} is taken into account naturally by the nonlinear
nature of the coupled equations (\ref{Eq.2}--\ref{Eq.6}).

Following the idea of Kogut \cite{Kogut.75},  
we approximate the gluonic part of the Hamiltonian (\ref{Eq.1}) by
a discretized form on a gauge lattice. 
In contrast to Ref. \cite{Kogut.75}, however, we represent the fermions
through point-like particles.  
This leads to a Hamiltonian which is represented as a sum of the 
following terms 
\begin{eqnarray}
\label{Eq.7}
{H} = {H}_{\rm part}\,+\,{H}_{\rm YM}^{\rm (lattice)}
\end{eqnarray}
where ${H}_{\rm part}$ contains the first two terms on the r.h.s.
of (\ref{Eq.1}) and ${H}_{\rm YM}^{\rm (lattice)}$
is defined as
\begin{eqnarray}
\label{Eq.8}
{H}_{\rm YM}^{\rm (lattice)}&=& -a^3 \sum_{x,k} {\rm Tr}
                            \Bigl\{\, {\cal E}_{x,k}{\cal E}_{x,k}\Bigr. 
\nonumber\\
                       &\,& -\Bigl( {1\over{4iga^2}} 
                            \sum_{l,m}\varepsilon_{klm}
                            ({\cal U}_{x,ml}-{\cal U}_{x,lm}) \Bigr)^2
\label{Eq.9} \\
                       &\,& -\Bigl. g\,{\cal J}_{x,k}\,{\cal A}_{x,k} \Bigr\}
\nonumber.
\end{eqnarray}

As already mentioned, the dynamical equations (\ref{Eq.2}--\ref{Eq.6}) 
can be solved efficiently by numerical time integration.  A lattice version 
of the continuum equations is constructed \cite{Hu.97,Moore.98} by 
expressing the gauge fields in terms of link variables 
${\cal U}_{x,l}\,\epsilon$ SU($N$), which represent the parallel transport 
of a field amplitude from a site $x$ to a neighboring site $(x+l)$ in 
the direction $l$.  As in the Kogut-Susskind model \cite{Kogut.75} we 
choose the temporal gauge $A_0 = 0$ and define the following variables:
\begin{eqnarray}
\label{Eq.11}
{\cal U}_{x,l} 
&=& \exp(-iga{\cal A}_l(x)) \,\,=\,\, {\cal U}^{\dagger}_{x+l,-l} \\
\label{Eq.12}
{\cal U}_{x,kl} 
&=& {\cal U}_{x,k}\, {\cal U}_{x+k,l}\, {\cal U}_{x+k+l,-k}\, 
                  {\cal U}_{x+l,-l} 
\end{eqnarray}
Consequently, we have
\begin{eqnarray}
\label{Eq.13}
{\cal E}_{x,j} &=& { 1\over{iga}}\,\dot {\cal U}_{x,j}\,
                   {\cal U}^{\dagger}_{x,j}  \\  
\label{Eq.14}
{\cal B}_{x,j} &=& { 1\over{4iga^2}}\,
                   \epsilon_{jkl}\, ({\cal U}^{\dagger}_{x,kl} - 
                  {\cal U}_{x,kl} )
\end{eqnarray}
for the electric and magnetic fields 
(${\cal E}_{x,j},{\cal B}_{x,j}\,\epsilon$ LSU($N$)), respectively. 
There are advantages in choosing 
${\cal U}_{x,i}$ and ${\cal E}_{x,i}$ as the 
basic dynamic field variables. This choice transforms the discretized
Yang-Mills equations into the following equations of motion
\begin{eqnarray}
\label{Eq.15}
\dot{\cal U}_{x,k}(t) &=& i\,g\,a\,{\cal E}_{x,k}(t)\,{\cal U}_{x,k}(t) \\
\label{Eq.16}
\dot{\cal E}_{x,k}(t) &=& {1\over{2iga^3}}\sum\limits^3_{l=1}
\Bigl\{ {\cal U}^{\dagger}_{x,kl}(t) -{\cal U}_{x,kl}(t)\,             \\
&-& {\cal U}^{\dagger}_{x-l,l}(t)\, {\cal U}^{\dagger}_{x-l,kl}(t)
{\cal U}_{x-l,l}(t)\, 
\nonumber \\
&+& {\cal U}^{\dagger}_{x-l,l}(t)\, {\cal U}_{x-l,kl}(t)\, {\cal U}_{x-l,l}(t)
\Bigr\}
\nonumber\\
&-& g\,{\cal J}_{x,k}(t). 
\nonumber 
\end{eqnarray}
In the spirit of the statistical nature of the transport theory, we split 
each quark into a number $n_q$ of test particles, each of which carries the 
fraction $q_0=Q_0/n_q$ of the quark color charge $Q_0$. In a first step, 
we adopt the gauge group SU(2) here for simplicity. 
Consequently, each nucleon is 
represented by two quarks (instead of three), initially carrying opposite 
color charge.

Perturbative short range interactions at high momenta can be described
in form of a stochastic collision term, well known from Boltzmann-type
transport equations \cite{Bertsch.84,Kruse.85}. 
For a consistent description of both,
long range and short range interactions on an equal footing, the equations 
of motion (\ref{Eq.3} -- \ref{Eq.5}) for the long range interactions 
have to be cast into the form of a single transport equation 
and combined with the collision term.
The Vlasov part of the transport equation, from which the equations of 
motion (\ref{Eq.3} -- \ref{Eq.5}) can be recovered, was first derived
in  \cite{Heinz.83,Elze.89}. We extend the formulation by adding a 
stochastic collision term similar to the one used in \cite{Geiger.92}.
The full transport equation then follows as:
\begin{eqnarray}
p_i^0 \, \frac{{\rm d f}_k (x_i^\mu,p_i^\mu,{\cal Q}_i)}{{\rm d} t} & \equiv &
p_i^\mu \, \left\{ \partial_\mu \,-\, 2g\,
	{\rm Tr}\bigl({\cal Q}_i {\cal F}^{\mu\nu}\bigr)\, \partial_p^\nu
        \right.  \nonumber \\
        & &   +2\,\imath g\, {\rm Tr}\bigl([{\cal Q}_i, 
	{\cal A}^{\mu}]_- \partial_{\cal Q}\bigr) \Bigr\} 
        {\rm f}_k(x_i^\mu,p_i^\mu,{\cal Q}_i) 
	\nonumber\\
        & = & \sum\limits_{\rm processes} C(p_i^\mu,x_i^\mu,{\cal Q}_i, t) .
\end{eqnarray}
Here f$_k$ denotes the one-particle distribution functions of the valence 
quarks and of the ``hard'' gluons ($k=$ q, g). This set of nonlinear 
integro-differential equations is coupled to the Yang-Mills equation in 
which the color current is now given by a moment of the one-particle 
distribution functions:
\begin{equation}
D_\mu {\cal F}^{\mu\nu}(x) \,=\, g\,
\sum\limits_k \int {\rm d}{\cal Q}_i \, \frac{{\rm d}^3 p_i}{p_i^0} \, 
        {\cal Q}_i \, p_i^\nu \, 
        {\rm f}_k(p_i^\nu,x_i^\nu,{\cal Q}_i)
\end{equation}

The collision integrals have the form:
\begin{eqnarray}
\label{ceq1}
C(p_i^\mu,x_i^\mu,{\cal Q}_i,\tau) &=& \frac{1}{2 S_i} \cdot
\int \theta(|p_i| - |k_c|)\,
\prod\limits_j {\rm d}\Gamma_j \, | {\cal M}^{(c)} |^2 
\nonumber \\
      & & (2 \pi)^4 \, \delta^4(P_{\rm in} - P_{\rm out}) \, 
	  D({\rm f}_k(p_i^\mu,x_i^\mu,{\cal Q}_i)) 
\end{eqnarray}
with
\begin{equation}
D({\rm f}_k(p_i^\nu,x_i^\nu,{\cal Q}_i)) \,=\, 
\prod\limits_{\rm in} {\rm f}_k(p_i^\nu,x_i^\nu,{\cal Q}_i) \, - \,
\prod\limits_{\rm out} {\rm f}_k(p_i^\nu,x_i^\nu,{\cal Q}_i) \quad,
\end{equation}
and
\begin{equation}
\prod\limits_j {\rm d}\Gamma_j = \prod\limits_{{j \ne i} \atop {\rm  in,out}} 
        \frac{{\rm d}^3 p_j}{(2\pi^3)(2p^0_j)} \, \theta(|p_j| - |k_c|) 
\quad.   
\end{equation}
$S_i$ is a statistical factor defined as
\begin{equation}
S_i \,=\, \prod\limits_{j \ne i} K_a^{\rm in}!\, K_a^{\rm out}!
\end{equation}
with $K_a^{\rm in,out}$ identical partons of species $a$ in the initial
or final state of the process, excluding the $i$th parton.

The step functions $\theta(|p_i| - |k_c|)$ insure that only hard
particles are allowed to propagate in the system. The superscript
$(c)$ on the matrix element ${\cal M}$ indicates that only the hard, 
i.e. short range, part of the interaction is treated in the collision term. 
This cut-off will be discussed in more detail below.

The matrix elements $| {\cal M}^{(c)} |^2$ account for the following 
processes:
\begin{eqnarray}
\label{processes}
{\rm A} \qquad & q + q' &\to\, q + q' \quad, \nonumber \\
{\rm B} \qquad & q + q  &\to\, q + q  \quad, \nonumber \\
{\rm C} \qquad & q + \bar q &\to\, g + g \quad, \nonumber \\
{\rm D} \qquad & g + g &\to\, g + g  \quad,  
\end{eqnarray}
together with those obtained from crossing relations ($q$ and $q'$ denote 
different quark flavors).
The amplitudes for these processes -- not taking the infrared lattice cut-off
$k_c$ into account -- have been calculated in refs. 
\cite{Cutler.78,Combridge.77,Bengtsson.84} for massless quarks and 
in refs. \cite{Combridge.79,Nason.88} for massive quarks. The 
corresponding scattering cross sections are expressed in terms
of spin- and color-averaged amplitudes $|{\cal M}^{(c)}(\hat s, \hat t, \hat u)|^2$:
\begin{equation}
\label{dsigmadt}
\frac{{\rm d}\hat \sigma^{(\rm A,B,C,D)}(\hat s, \hat t, \hat u)}
     {{\rm d}\hat t} \,=\, \frac{1}{16 \pi \hat s^2}
	\,\langle |{\cal M}^{(c)}(\hat s, \hat t, \hat u)|^2 \rangle
\end{equation}
with $\hat s, \hat t, \hat u$ being the well-known Mandelstam variables.
For the transport calculation we also need the total cross section 
as a function of $\hat s$ which can be obtained from (\ref{dsigmadt}):
\begin{equation}
\label{sigmatot}
\hat \sigma_{ab}(\hat s) \,=\, 
\sum\limits_{c,d} \, \int\limits_{\hat t_{\rm min}}^{\hat t_{\rm max}}
	\left( \frac{{\rm d}\hat \sigma
	(\hat s, \hat t', \hat u)}{{\rm d}\hat t'}
	\right)_{ab\to cd} {\rm d}\hat t' \quad .
\end{equation}
The integration boundaries are fixed through kinematical constraints.
Note that the treatment of the cross section (\ref{processes}--\ref{sigmatot})
with the matrix elements supplied in 
\cite{Cutler.78,Combridge.77,Bengtsson.84,Combridge.79,Nason.88}
does not take the infrared lattice cut-off $k_c$ into account.
The rigorous way to evaluate the matrix elements $| {\cal M}^{(c)} |^2$
and to eliminate the small momenta from the gluon propagators 
would be to subtract the lattice propagator from the continuum propagator
in the Feynman diagram describing the scattering process at lowest order.
Because the evaluation of the gluon propagator on the lattice is complicated, 
we propose here to use, for exploratory studies, the usual matrix elements 
\cite{Cutler.78,Combridge.77,Bengtsson.84,Combridge.79,Nason.88} but with
a cut-off on the allowed momentum transfer, corresponding to the lattice
cut-off $k_c = \pi/a$. We can cast this into the Lorentz invariant form
that the scale of the interaction, $Q^2(\hat s, \hat t, \hat u)$
must satisfy the constraint 
\begin{equation}
\label{constraint}
Q^2(\hat s, \hat t, \hat u) > k_c^2 \quad.
\end{equation}
The functional form of $Q^2$ is generally process dependent and not
unambiguous, although at leading order all choices for $Q$ that increase
with the parton-parton center-of-mass energy are equivalent.
One can now solve equation (\ref{constraint}) for $\hat t$ in order
to obtain an additional constraint for the integration boundaries
of equation (\ref{sigmatot}). Thus, only momentum transfers larger
than $k_c$ contribute to the total cross section.
It was shown in ref. \cite{Gyulassy.97} that spectrum of the classical 
Yang-Mills radiation matches smoothly onto the conventional minijet
distribution near the intrinsic transverse momentum scale of the partons 
in a heavy nucleus at high energy. The resulting expectation that the
precise choice of the momentum cutoff $k_c$ is not important must, of
course, be verified by future numerical calculations.

In summary, we have developed a novel formalism,
which allows for the first time the treatment of both, 
the hard and the soft parton dynamics in ultrarelativistic heavy ion 
collisions in a consistent transport approach: 
The emission of gluons from the initially coherent parton 
configurations of the colliding nuclei as well as 
low-$p_t$ color coherence effects in parton-parton scatterings are
treated in the framework of a Yang-Mills transport equation on a coupled
lattice-particle system. Intermediate and high-$p_T$ interactions
are described in a collision term similar to that of the parton
cascade model. This formalism thus avoids problems connected to the
infrared cut-offs in the parton cascade model and offers
a unified treatment of coherence effects within that approach.

{\it Acknowledgments:} We gratefully acknowledge remarks from Ulrich
Heinz which helped to improve our manuscript.
One of us (S.A.B.) acknowledges support from a
Feodor Lynen Fellowship of the Alexander v. Humboldt Foundation.
This work was supported in part by a grant from the U.S. Department
of Energy DE-FG02-96ER40495.



\end{document}